\documentstyle[aps,prd,12pt,tighten,amssymb,eqsecnum,axodraw,cite]{revtex}

\title{\vspace*{1.5in}Light--Cone Broadening and TeV Scale Extra
Dimensions}

\author{{\bf A. Campbell--Smith}$^{a,b}$, {\bf John Ellis}$^b$, {\bf N.E.
  Mavromatos}$^{a,b}$ and {\bf D.V. Nanopoulos}$^{c}$}

\address{\vspace*{0.5cm}${}^a$ Theoretical Physics (University of Oxford), 1 Keble
  Road, OXFORD, OX1 3NP, U.K.,\\${}^b$ Theory Division, CERN, CH--1211
  Geneva 23, Switzerland,\\${}^c$Department of Physics, Texas A \& M
  University, College Station, TX~77843--4242, USA, Astroparticle
  Physics Group, Houston Advanced Research Center (HARC), Mitchell
  Campus, Woodlands, TX 77381, USA, and Academy of Athens, Chair of
  Theoretical Physics, Division of Natural Sciences, 28~Panepistimiou
  Avenue, Athens 10679, Greece.}

\newcommand{\bard}{d \!\!\rule[6.5pt]{4pt}{0.3pt}}

\begin{document}

\maketitle
\vspace*{0.4cm}
\begin{center}
{\bf Abstract}
\end{center}
\vspace*{0.4cm}
\begin{abstract}
  We examine the effect of light--cone broadening induced by
  quantum--gravity foam in the context of theories with ``large'' extra
  dimensions stretching between two parallel brane worlds.  We
  consider the propagation of photon probes on one of the branes,
  including the response to graviton fluctuations, from both field--
  and string--theoretical viewpoints. In the latter approach, the
  dominant source of light--cone broadening may be the recoil of the
  D--brane, which scales linearly with the string coupling.
  Astrophysical constraints then place strong restrictions on
  consistent string models of macroscopic extra dimensions.  The
  broadening we find in the field--theoretical picture seems to be
  close to the current sensitivity of gravity--wave interferometers,
  and therefore could perhaps be tested experimentally in the
  foreseeable future.
\end{abstract}

\vspace*{-6.5in}
\begin{flushright}
  CERN--TH/99--207\\
  OUTP--99--33P\\
  hep-th/9907141
\end{flushright}
\vspace*{5.5in}

\section{Introduction}
\label{sec:intro}

The considerable theoretical interest in the recent
suggestion~\cite{dimopo98} of possible extra dimensions of macroscopic
(sub--millimetre) size stems largely from the apparent extreme
difficulty of ruling out such a suggestion experimentally.  It is the
point of this note to examine this suggestion from the point of view
that quantum gravity might be treated as a stochastic medium
\cite{ellis84,ellis92,garay98,ashtekar99}.  As three of us have argued
in the past~\cite{ellis98}, when one considers interactions of
closed--string particle states with D--branes, the recoil of the
latter may induce distortions of the space--time around the brane,
which manifest themselves as new quantum degrees of freedom, carrying
information and giving a stochastic nature to the process.  One of the
most important features of this approach is the induced light--cone
fluctuations~\cite{ellis99}, which may be detectable as stochastic
fluctuations in the velocity of light propagating through this
``medium''.

Such phenomena may also characterize conventional point--like
approaches to quantum gravity. Indeed, as previously argued
in~\cite{ford95}, one encounters light--cone fluctuations when one
expands about a squeezed graviton coherent state, which arguably
characterizes physically interesting models of quantum--gravity foam.
The application of such ideas to flat Minkowskian space--time with
compactified extra dimensions was first considered
in~\cite{ford99,ford99_2}, with the conclusion that the stochastic
fluctuations in the light--cone produce stochastic fluctuations in the
arrival times of photons which are considerably larger than the
conventional Planck scale of four--dimensional gravity.
In~\cite{ford99_2}, a similar calculation has been performed in the
context of the models of~\cite{dimopo98}, but the calculation has been
done only in the case of just one extra dimension, which may be ruled
out by macroscopic astrophysical observations~\cite{dimopbig99}.

\begin{figure}
\begin{center}
\begin{picture}(155,155)(0,-5)
\Line(10,10)(10,110)
\Line(10,10)(45,45)
\Line(10,110)(45,145)
\Line(45,45)(45,145)
\Line(110,10)(110,110)
\Line(110,10)(145,45)
\Line(110,110)(145,145)
\Line(145,45)(145,145)
\LongArrow(60,5)(110,5)
\LongArrow(60,5)(10,5)
\Text(60,12)[c]{$l=\pi\Lambda$}
\Text(12,33)[l]{$D_1$}
\Text(112,33)[l]{$D_2$}
\PhotonArc(75,105)(12,0,360){1.7}{9}
\LongArrow(80,105)(100,105)
\DashLine(80,105)(63,105){2}
\Line(63,105)(50,105)
\DashArrowLine(28,33)(28,123){5}
\PhotonArc(28,78)(12,270,90){1.7}{4.5}
\Vertex(28,90){1}
\Vertex(28,66){1}
\DashLine(28,118)(28,150){5}
\DashLine(18,118)(18,150){2}
\Text(23,140)[c]{$\ell_p$}
\end{picture}
\caption{Schematic representation of the geometry of interest for our
  computation, where $D_1$ and $D_2$ are brane worlds separated by the
  size of the extra dimensions $l$, in which only closed--string states
  (gravitons) can propagate. A photon is depicted travelling
  parallel to $D_1$ and separated from it by a small distance
  $\ell_p$, reflecting the quantum uncertainty in the location of the
  brane.\label{fig1}}
\end{center}
\end{figure}
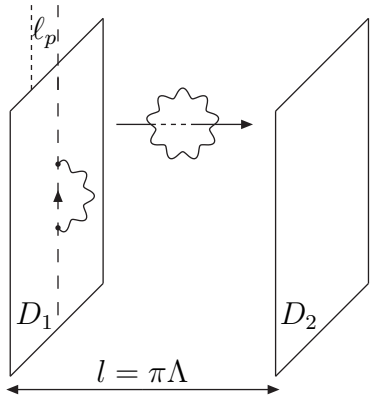

In the present work, we consider the stochastic fluctuations of the
light--cone in a rather different framework, that of two parallel
3--branes separated by a distance $l$.  The branes live in a world
with $n \leqslant 6$ extra transverse dimensions, in the standard string
theory picture. As discussed in~\cite{dimopo98}, only closed--string
states (gravitons) can propagate in the bulk.  The geometry of
interest is depicted in Fig.~\ref{fig1}.  We initially ignore the
recoil of the branes, and only study the effect of graviton
fluctuations about flat $(3\!+\!1)$--dimensional space--time on the
brane, which is assumed to have Minkowskian signature
$({+}{-}{-}{-})$.  The issue of recoil effects is taken up later.

In the picture of~\cite{dimopo98}, the Planck scale on the branes is
constrained to be the standard Planck mass scale $ M_p^{(4)} \sim
10^{19}$~GeV, which is related to the underlying
$(4\!+\!n)$--dimensional scale $M_p^{(4+n)}$ via the size $l=\pi\Lambda$
of the postulated extra dimensions:
\begin{eqnarray}
  \label{planckscales}
    \left(M_p^{(4)}\right)^2 \sim \left(\pi\Lambda\right)^n
  \left(M_p^{(4+n)}\right)^{2+n}.
\end{eqnarray}
We are interested in the superstring--motivated cases of $n\!
\leqslant \!6$ extra dimensions, and we work in units such that
\(\hbar = c = M_p^{(4+n)} = 1\).

We first review briefly the analysis in~\cite{ford95,ford99}, which
considered gravitons in a squeezed coherent state, relevant for
discussions of quantum--gravitational space--time foam.  Such
gravitons induce quantum fluctuations in the space--time metric, in
particular fluctuations in the light--cone, which may have observable
effects $\Delta t$ on the arrival times of photons.  Consider a flat
background space--time with a linearized perturbation corresponding to
the invariant metric element
\[ds^2=g_{\mu\nu}dx^\mu dx^\nu = 
\left(\eta_{\mu\nu} + h_{\mu\nu}\right)dx^\mu dx^\nu = dt^2 -
d\vec{x}^2 + h_{\mu\nu}dx^\mu dx^\nu.\] Let $2 \sigma (x,x')$ be the
squared geodesic separation for any pair of space--time points $x$ and
$x'$, and let $2 \sigma_0(x,x')$ denote the corresponding quantity in
an unperturbed flat space--time background.  In the case of small
gravitational perturbations about the flat background, one may expand
$\sigma = \sigma _0 + \sigma_1 + \sigma_2 + \dots$, where $\sigma_n$
denotes the $n^{\mathrm{th}}$--order term in an expansion in the
gravitational perturbation $h_{\mu\nu}$.  Then, as shown
in~\cite{ford95}, the root--mean--square deviation from the classical
propagation time $\Delta t$ is related to $\langle\sigma^2\rangle$ by:
\begin{eqnarray}
\label{dtdef}
\Delta t = \frac{\sqrt{\langle\sigma^2\rangle -
\langle\sigma_0^2\rangle}}{L} \simeq \frac{\sqrt{\langle\sigma_1^2\rangle}}{L} + \dots
\end{eqnarray}
where $L = |x' - x|$ is the distance between the source and the
detector.  The expression (\ref{dtdef}) is gauge
invariant~\cite{ford99}.  For convenience, the transverse trace--free
gauge is used, for which \(h_{0\nu}=h_i^i=\partial^\mu h_{\nu\mu}=0,\)
where Greek indices refer to \((3\!+\!n\!+\!1)\)--dimensional space--time
and Latin indices refer to the \(3\!+\!n\) spatial components only.

The light--cone broadening effect is computed in~\cite{ford99} both
for a cylindrical topology (or one periodic compactified dimension)
and in the presence of a plane boundary.  The situation of most
interest for us is that in which the photon travels in a direction
orthogonal to the compactified dimension, i.e., parallel to the plane
boundary, since only closed--string states (gravitons) propagate in
the extra dimensions~\cite{dimopo98}).  The following results were
derived in~\cite{ford99}:
\begin{eqnarray}
  \label{fordcyl}
  \frac{\Delta t}{t_p} \simeq \frac{1}{4\surd 2} \sqrt{\frac{L}{l}},
\end{eqnarray}
for the cylindrical topology, where $l$ is the size of the compact
dimension, and $L\gg l$ is the distance travelled by the photon, and
in the vicinity of a plane boundary (at which the gravitons were
forced to obey Neumann boundary conditions):
\begin{eqnarray}
  \frac{\Delta t}{t_p} \simeq \sqrt{\frac{\ln[L/z]}{6\pi^2}},
\end{eqnarray}
where $z$ is the distance of the photon trajectory from the plane, for
which it is assumed that $z\ll L$.  The technique used in the above
computations was that of the image method, in which the Kaluza--Klein
modes resulting from the compact dimension are taken into account in
the graviton two--point function
\[G_{ijkl}(x,t;x',t') = \langle h_{ij}(x,t)\,h_{kl}(x',t')\rangle\] as
follows~\cite{ford99}:
\begin{eqnarray}
  \label{imagesum}
  G_{xxxx}(t,\vec{x},z_i;t',\vec{x}',z_i') =
  {\sum_{m_i=-\infty}^{\infty}}'
  G_{xxxx}(t,\vec{x},z_i;t',\vec{x}',z_i'+m_i l), 
\end{eqnarray}
where $z_i$ label the compact dimensions and the prime on the sums
indicates that the $m_i=0$ terms are omitted.  This technique
simplifies considerably the calculation of the contribution from the
Kaluza--Klein modes for the case where the compactified dimension is
periodic.

A similar computation has been performed in~\cite{ford99_2}, this time
concentrating on a cylindrical topology in more than four dimensions,
i.e., with some extra compact dimensions, as motivated by the proposal
of~\cite{dimopo98}. It was found that, for a five--dimensional world
with one dimension compact, the light--cone broadening is
\begin{eqnarray}
  \Delta t \sim t_p \, \frac{L}{l},
\end{eqnarray}
where $t_p$ is the Planck time.  Note that this differs from the
cylindrical case (\ref{fordcyl}) above~\cite{ford99}, for which the
light--cone broadening grows as the square root of $L$.  The
conclusion of~\cite{ford99_2} is that this scaling will in general
vary according to the dimensionality of the brane.

\section{Field--Theoretical Contribution}

Here we deal with a different topology (see Fig.~\ref{fig1}), in which
the extra dimensions do not have periodic boundary conditions.
Instead, we adopt Neumann boundary conditions for the graviton field
at each brane hyperplane.  Therefore we do not use the image method,
but instead obtain an estimate of the induced light--cone broadening
by identifying the dominant contributions in the various discrete mode
sums involved, which will be sufficient for our purposes. The
light--cone broadening effect is given by (\ref{dtdef}), with $L$
denoting the distance travelled by the photon and where \(\sigma_1\)
is related to the graviton two--point function by~\cite{ford99}
\begin{eqnarray}
  \label{sigmadef}
  \langle \sigma_1^2\rangle &=& \frac{1}{8} L^2 \int_{r_i}^{r_f} dr
  \int_{r_i}^{r_f} dr' \; n^a n^b n^c n^d \; \langle h_{ab} (x)
  h_{cd}(x') + h_{ab} (x') h_{cd}(x)\rangle \nonumber\\ 
  &\doteq& \frac{1}{8} L^2 \int_{r_i}^{r_f} dr
  \int_{r_i}^{r_f} dr' \; n^a n^b n^c n^d \; G_{abcd}(x,x'),
\end{eqnarray}
where the integrations are taken along the geodesic, the spatial
direction of which is defined by the spatial unit vectors $n$.

As already mentioned, the topology we wish to consider in evaluating
$\sigma_1$ is that of $n$ extra dimensions of size $l=\pi\Lambda$
\cite{dimopo98}, as shown in Fig.~\ref{fig1}.  We compute the effect
of this topology on the arrival time of a photon travelling parallel
to one of the boundaries but separated from it by a distance
$z\sim\ell_p$.  This last feature is supposed to model the effect of
quantum fluctuations in the position of the brane
world~\cite{mavro+szabo}.  This uncertainty is motivated by the
framework of general relativity, in which there cannot be rigid
planes.

Our conventions are as follows:
\begin{eqnarray}
  x^\mu &\doteq& ( t,\vec{x},z_1, \ldots, z_n) \nonumber\\
  k^\mu &\doteq& ( \omega, \vec{k}, m_1/\Lambda, \ldots, m_n/\Lambda )\; :
  \qquad m_i \in {\mathbb{Z}},
\end{eqnarray}
and, on account of the $n$ extra dimensions, the graviton modes are
normalized with an extra factor of $l^{-n/2}=(\pi\Lambda)^{-1/2}$.
Following~\cite{ford99}, the required two--point function can be
expressed as follows:
\begin{eqnarray}
  \label{2ptfn}
  G_{xxxx}(x,x') = 2\left( D(x,x') - 2 F_{xx} (x,x') + H_{xxxx} \right),
\end{eqnarray}
where
\begin{eqnarray}
  D(x,x') = -\frac{1}{4\pi^2} \left[ (t-t')^2 - (\vec{x}-\vec{x}')^2
  \right]^{-1}
\end{eqnarray}
is the Hadamard function for a free scalar field, and the functions $F$
and $H$ are defined by
\begin{eqnarray}
  F_{ij} &\doteq& {\mathtt{R\!e}} \prod_{a=1}^{n} \left[ \frac{1}{l}
  \sum_{m_a=-\infty}^{\infty} \right] \int \bard^3 k\; \frac{k_i
  k_j}{2\omega^3} e^{i k\cdot (x-x')} e^{-i \omega (t-t')},\\
  H_{ijkl} &\doteq& {\mathtt{R\!e}}\prod_{a=1}^{n} \left[ \frac{1}{l}
  \sum_{m_a=-\infty}^{\infty} \right] \int \bard^3 k \; \frac{k_i k_j k_k
  k_l}{2 \omega^5} e^{i k\cdot (x-x')} e^{-i \omega (t-t')}.
\end{eqnarray}
We have restricted ourselves above to consideration of the $xxxx$
component of the two--point function, since we are interested in
photons propagating parallel to the brane, and we conveniently set
this direction to be parallel to the $x$--axis, assuming rotational
invariance on the brane.  Adopting Neumann boundary conditions for the
graviton modes at the boundaries of each compact dimension, the
required function $F_{xx}$ is given by
\begin{eqnarray}
  F_{xx} = {\mathtt{R\!e}} \frac{\partial_x
  \partial_x'}{\pi^n\Lambda^n} \left[ \prod_{a=1}^{n}
  \sum_{m_a=-\infty}^{\infty} \cos\left[ m_i (z_i -
  z_i')/\Lambda \right] \right] \int \bard^3 k\;
  \frac{1}{2\omega^3} e^{i \vec{k} \cdot (\vec{x} - \vec{x}') -i
  \omega (t-t')},
\end{eqnarray}
where as a result of the transverse trace--free gauge choice~\cite{ford99}
\[ \omega = \sqrt{ k^2 + \alpha_n^2} ; \qquad \alpha_b^2
\doteq \frac{1}{\Lambda^2} \sum_{a=1}^b m_a^2 \doteq
\frac{1}{\Lambda^2} \overline{\alpha}_b^2.\] Performing the angular
part of the integration, we obtain
\begin{eqnarray}
  \label{fint}
  F_{xx} &=& \sum_{b=0}^{n} \frac{1}{4\pi^2} \frac{1}{\pi^n\Lambda^n}
  \partial_x \partial_x' \prod_{a=1}^{n-b} \left[ 2
  \sum_{m_a=1}^\infty \cos\left[ m_a Z_a/\Lambda \right]
  \right] \times \nonumber\\
  &&\qquad\times\int dk\; \frac{k}{\left( k^2 + \alpha_{n-b}^2\right)^{3/2}}
  \frac{\sin[kR]}{R} \cos[ (k^2+ \alpha^2_{n-b})^{1/2} T],
\end{eqnarray}
where we have written $T\doteq (t-t')$, $R\doteq |x-x'|$, and $Z_i
\doteq z_i - z_i'$.  In the sum over $b$, the $b^{\mathrm{th}}$ term
has $b$ of the mode numbers $m_a$ vanishing.

The term with $b=n$ has no sums to evaluate, and when combined with
the equivalent term from $H$ and the term $D$ as in~(\ref{2ptfn}), it
reduces to the expression computed in~\cite{ford99} for a single plane
boundary, and ultimately gives~\cite{ford99}
\begin{eqnarray}
  \label{zeromode}
  \Delta t = \frac{1}{(\pi\Lambda)^n} \sqrt{\frac{\ln \left[ L/\ell_p\right]}{6\pi^2}},
\end{eqnarray}
where $L$ is the distance travelled by the photon.  

We wish now to determine the effects of the non--zero modes in the sum
above, i.e., those which are sensitive to the presence of the other
plane boundaries.  Note first from~(\ref{fint}) that the effects of
high modes $m_a\gg 1$ are (as expected) small compared to the lowest
few modes with $m_a = 1,2,\ldots$.  Note also from (\ref{fint}) that,
after differentiation with respect to $x$ and $x'$, there will be some
terms after the integrations which will feel the presence of the
ultraviolet cutoff. In this way, we will be able to make a direct
connection with the relation between the four--dimensional Planck
scale and the underlying TeV scale of gravity~\cite{dimopo98}, as
shown in (\ref{planckscales}). In order to obtain an estimate of the
effects of the first non--zero modes in each of the extra dimensions,
we compute only the $m_a =1$ terms in the sums above.  The pertinent
expression is as follows:
\begin{eqnarray}
  &&\frac{d^2}{dR^2} \int dk\; \frac{k}{\left( k^2 +
  \Lambda^{-2}\right)^{3/2}} \frac{\sin[kR]}{R} \;\cos\left[ \left
  ( k^2 + \Lambda^{-2}\right)^{1/2} T \right] \nonumber\\
  &=&\frac{1}{\Lambda^2} \frac{d^2}{d\rho^2} \int dy\; \frac{y}{\left
  ( y^2 + \rho^2\right)^{3/2}} \sin[y] \cos\left[ \left( y^2 +
  \rho^2\right)^{1/2} \frac{T}{R}\right],
\end{eqnarray}
where $\rho\doteq R/\Lambda$ and $y\doteq kR$.  In order to estimate
the magnitude of this contribution, we split the integral at $\rho$
and approximate the kernel appropriately in each domain:
\begin{eqnarray}
  \frac{1}{\Lambda^2} \frac{d^2}{d\rho^2} \left\{ \int_0^\rho dy\;
  \frac{y\sin[y]\cos[t/\Lambda]}{\rho^3} + \int_\rho^C dy\;
  \frac{\sin[y]\cos[yT/\Lambda\rho]}{y^2} \right\},
\end{eqnarray}
where $C\sim M_p^{(4)} R$ is the ultraviolet cutoff, which can be
related via (\ref{planckscales}) to the size of the compact
dimensions.  The differentiations can be performed with ease, and
neglecting the trigonometric dependence one obtains an
order--of--magnitude estimate of the leading contribution for \(cT\gg
\Lambda\):
\begin{eqnarray}
  &&\frac{1}{\Lambda^2} \int_\rho^C dy\;
  \frac{T^2}{\Lambda^2\rho^4} \sim \frac{M_p^{(4)}}{R} \sim
  \frac{(\pi\Lambda)^{n/2}}{R},
\end{eqnarray}
where we have adopted the approximation $R=T$ in the kernel,
consistent with the classical (unbroadened) light--cone.  Now we can
reassemble the function $F$ and integrate over $x$ and $x'$ as in
(\ref{sigmadef}), which yields a light--cone broadening of the form
\begin{eqnarray}
  \frac{\Delta t}{t_p} \sim \frac{2^{n/2}}{2\pi}
  \left(\frac{L}{\ell_p}\right)^{1/2} \sqrt{\ln [L/\ell_p]}\;
  \left(\frac{\ell_p}{\pi\Lambda}\right)^{n/4},
\label{twothirteen}
\end{eqnarray}
which for the case $n\!=\!6$ reduces to:
\begin{eqnarray}
  \frac{\Delta t}{t_p} \sim \frac{4}{\pi^{5/2}}
  \left(\frac{L}{\Lambda}\right)^{1/2}
  \left(\frac{\ell_p}{\Lambda}\right) \sqrt{\ln[L/\ell_p]}.
\label{twofourteen}
\end{eqnarray}
We have neglected contributions from the function $H$, since they die
away much more quickly with increasing $m_i$ and will not affect our
rough estimate.  Note that in the expression above we have not removed
the factor of \((\pi\Lambda)^{-3}\) from the graviton normalization,
which is common to the zero--mode (\ref{zeromode}), this contribution,
and also the vacuum contribution.  In a string--theoretical picture,
this scale $\Lambda$ could pick up a dynamical significance, and could
vary with time. As was noted in~\cite{ford99_2}, time variation of
\(\Lambda\) could also result from cosmological expansion.  Note
finally that the $L$ dependence is very similar to that obtained
in~\cite{ford99} for a compactified topology, as seen in
(\ref{fordcyl}) above.

The ratio $\ell_p/\pi\Lambda$ can be calculated in terms of the
fundamental Planck scale $M_p^{(4+n)}$, as follows.
Following~\cite{dimopo98}, we have
\begin{eqnarray}
  \ell_p &=& 2\times 10^{-17} \left(\frac{1 {\mathrm{TeV}}}{M_p^{(4+n)}}
  \right) {\mathrm cm}\nonumber\\
  \pi\Lambda &=& 10^{30/n - 17} \left(\frac{1 {\mathrm{TeV}}}{M_p^{(4+n)}}
  \right)^{1+2/n} {\mathrm cm}
\end{eqnarray}
whence, if $M_p^{(4+n)}= \mu \, {\mathrm{TeV}}$,
\begin{eqnarray}
  \left( \frac{\ell_p}{\pi\Lambda}\right) = 2 \times 10^{-30/n} \mu^{2/n}.
\end{eqnarray}
For $n=6$, this ratio is bigger than \(2\times 10^{-5}\) and grows with
$\mu$ as \(\mu^{1/3}\), and for $\mu\sim 100$ it can be as large as
$10^{-4}$.

To gain an understanding whether the light--cone broadening effect
might in principle be measurable, we consider some specific cases.
For gamma--ray bursters with redshifts $z \sim 1$, $L\sim 10^{28}
\mathrm{cm}$, and, using the above formulae, we estimate $\Lambda\sim
10^{-12} \mathrm{cm}$, $\ell_p\sim 10^{-17} \mathrm{cm}$ for $ n = 6$,
and hence
\begin{eqnarray}
  \Delta t \sim 10^{15} t_p,
\end{eqnarray}
In the extra dimension picture the (fundamental) Planck time is very
much larger than normal, being of order $10^{-27}$ seconds, so that
the light--cone broadening is \(10^{-12}\) seconds.  This is far below
the sensitivity of experiments measuring gamma--ray
bursts~\cite{amelino98}, which is in the millisecond region.  It is
easy to see that the effect is even smaller for $ n < 6$.  The above
estimates have been made for $\mu\!=\!1$; if the underlying scale is
significantly higher than this, both estimates would get larger.  In
the case of gravity--wave interferometers, the sensitivity of the
experiments is much better~\cite{amelino99}, namely of the order of
$10^{-18}$ metres.  For this case we have $L\sim 10^{3} \mathrm{cm}$,
and
\begin{eqnarray}
  \Delta t \sim 10^2 t_p \sim 10^{-25} \,\mathrm{s},
\end{eqnarray}
which is in principle testable at current or future gravity--wave
interferometers, provided there is a controlled way to
distinguish this effect from conventional noise sources.

Before closing this section, we would like to remark briefly on the
computation of~\cite{ford99_2} concerning extra compact dimensions.
Due to the periodic boundary conditions imposed in that case, there
are Kaluza--Klein modes which are resummed using the image method.
The case of more than one extra dimension complicates the analysis and
was not considered in detail in~\cite{ford99_2}.  For one extra
dimension, the light--cone broadening effect was found to scale
linearly with the distance travelled by the photon, in contrast to our
estimate (\ref{twothirteen}).  We stress therefore that this scaling
depends crucially on the boundary conditions as well as the number of
non--compact dimensions.  For example, when the image method is used
for the case of a periodic fifth dimension, there are four $k$
integrals in the non--compact space, and the effect of the
discretization is incorporated with the image sum (\ref{imagesum}).
In the case considered above, with only Neumann boundary conditions,
there are three $k$ integrations and the sums have to be performed
explicitly.

\section{Non--Critical String and the D--brane Recoil Contribution}

So far we have ignored recoil of the 3--brane, which is present in all
realistic cases due to the scattering with the closed--string state
that propagates in the bulk.  As discussed in~\cite{ellis98} in the
case of D--brane string solitons, their recoil after interaction with
a closed--string graviton state~\cite{kogan96} is characterized in a
world--sheet context by a $\sigma$--model deformed by pairs of
logarithmic operators~\cite{lcft}:
\begin{equation}
C^I_\epsilon \sim \epsilon \Theta_\epsilon (X^I),\qquad
D^I_\epsilon \sim X^I \Theta_\epsilon (X^I), \qquad I =0,\dots, 3 
\label{logpair}
\end{equation} 
defined on the boundary $\partial \Sigma$ of the string
world--sheet. Here $X^I$ obey Neumann boundary conditions 
on the string world--sheet, and denote the brane coordinates.
The remaining $y^i, i=4, \dots, 9$ denote the transverse directions.  

In the case of D--particles, examined in~\cite{ellis98}, $I$ takes the
value $0$ only.  In such a case, the operators (\ref{logpair}) act as
deformations of the conformal field theory on the world--sheet: $U_i
\int _{\partial \Sigma} \partial_n X^i D_\epsilon $ describes the
shift of the D--brane induced by the scattering, where $U_i$ is its
recoil velocity, and $ Y_i \int _{\partial \Sigma} \partial_n X^i
C_\epsilon$ describes quantum fluctuations in the initial position
$Y_i$ of the D--particle. It has been shown~\cite{mavro+szabo} that
energy--momentum is conserved during the recoil process: $U_i = k_1 -
k_2$, where $k_1 (k_2)$ is the momentum of the propagating
closed--string state before (after) the recoil, as a result of the
summation over world--sheet genera.  We also note that $U_i = g_s
P_i$, where $P_i$ is the momentum and $g_s$ is the string coupling,
which is assumed here to be weak enough to ensure that D--branes are
very massive, with mass $M_D=1/(\ell _s g_s)$, where $\ell _s$ is the
string length.

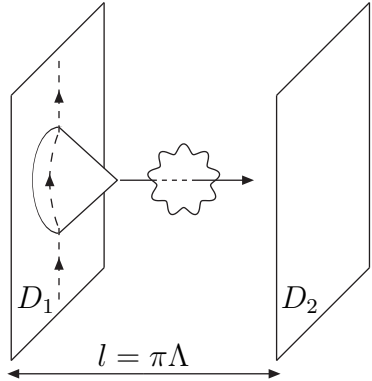
\begin{figure}
\begin{center}
\begin{picture}(155,155)(0,-5)
\Oval(28,78)(20,10)(0)
\CBox(28,58)(38,98){White}{White}
\Line(28,58)(50,78)
\Line(28,98)(50,78)
\Vertex(49.8,78){.3}
\Line(10,10)(10,110)
\Line(10,10)(45,45)
\Line(10,110)(45,145)
\Line(45,45)(45,73.5)
\Line(45,82.5)(45,145)
\Line(110,10)(110,110)
\Line(110,10)(145,45)
\Line(110,110)(145,145)
\Line(145,45)(145,145)
\LongArrow(60,5)(110,5)
\LongArrow(60,5)(10,5)
\Text(60,12)[c]{$l=\pi\Lambda$}
\Text(12,33)[l]{$D_1$}
\Text(112,33)[l]{$D_2$}
\PhotonArc(75,78)(12,0,360){1.7}{9}
\LongArrow(80,78)(100,78)
\DashLine(80,78)(63,78){2}
\Line(63,78)(51,78)
\DashArrowLine(28,33)(28,60){3}
\DashArrowLine(28,96)(28,123){3}
\DashArrowArcn(74.5,78)(50,200,160){3}
\end{picture}
\caption{Schematic representation of the recoil effect:  the photon's
  trajectory (dashed line) is distorted by the conical singularity in
  the brane that results from closed--string emission into the
  bulk.\label{fig2}}
\end{center}
\end{figure}

In the case of D--$p$--branes, the pertinent deformations are slightly
more complicated. As discussed in~\cite{kogan96}, the deformations are
given by $\sum_{I} g^1_{iI} \int _{\partial \Sigma} \partial_n X^i
D_\epsilon ^I$ and $ \sum_{I} g^2_{Ii} \int _{\partial \Sigma}
\partial_n X^i C^I_\epsilon $.  The $0i$ component of the `tensor'
couplings $g^{\alpha}_{Ii},~\alpha=1,2$ include the collective momenta
and coordinates of the D--brane, as before, but now there are more
couplings $g^{\alpha}_{Ii},~I\ne0$, describing the `bending' of the
D--brane under the emission of a closed--string state propagating in
the transverse direction, as seen in Fig.~\ref{fig2}.

The correct specification of the logarithmic pair (\ref{logpair})
entails a regulating parameter $\epsilon \rightarrow 0^+$, which
appears inside the $\Theta_\epsilon (t)$ operator: $\Theta_\epsilon
(X^I) = \int \frac{d\omega}{2\pi}\frac{1}{\omega -i\epsilon}
e^{i\omega X^I} $. In order to realize the logarithmic algebra between
the operators $C$ and $D$, one takes~\cite{kogan96}: $\epsilon^{-2}
\sim \ln [\Lambda/a] \doteq \beta$, where $\Lambda$ ($a$) are
infrared (ultraviolet) world--sheet cutoffs.  The recoil operators
(\ref{logpair}) are slightly relevant, in the sense of the
renormalization group for the world--sheet field theory, with small
conformal dimensions $\Delta _\epsilon = -\frac{\epsilon^2}{2}$.

The relevant two--point functions have the following form:
\begin{eqnarray}
&~&\langle C_\epsilon (z)C_\epsilon (0)\rangle \stackrel{\epsilon\to 0}{\sim} 0+
{\cal O}(\epsilon^2) \nonumber \\
&~&\langle C_\epsilon (z)D_\epsilon (0)\rangle 
\stackrel{\epsilon\to 0}{\sim}
{\pi\over2}\sqrt{{\pi\over\epsilon^2\beta}}
\left(1+2\epsilon^2 \log|z/a|^2 \right) \nonumber \\
&~&\langle D_\epsilon (z)D_\epsilon (0)\rangle =\frac{1}{\epsilon^2}
\langle C_\epsilon (z)D_\epsilon (0)\rangle \stackrel{\epsilon\to 0}{\sim}
{\pi\over2}\sqrt{{\pi\over\epsilon^2\beta}}
\left({1\over\epsilon^2}-2\eta\log|z/a|^2\right)
\label{twopoint}
\end{eqnarray} 
which is the logarithmic algebra~\cite{lcft} in the limit $\epsilon
\rightarrow 0^+$, modulo the leading divergence in the $\langle
D_\epsilon D_\epsilon\rangle$ recoil correlator.  In fact, it is this
leading divergent term that will be of importance for our purposes
below.

Since the recoil operators are relevant in a world--sheet
renormalization--group sense, they require dressing with a Liouville
field~\cite{ddk} in order to restore conformal invariance, which has
been lost in the recoil process. One then makes the crucial step of
identifying the world--sheet zero mode of the Liouville field with the
target time $t$, which is justified in~\cite{ellis98,mavro+szabo}
using the logarithmic algebra (\ref{twopoint}) for the case at hand.
This identification leads to the appearance of a curved space--time
background, with metric elements that generalize straightforwardly
those for D--particles, that were given in~\cite{kanti}:
\begin{equation} 
G_{ij} =\delta _{ij}, \quad G_{00}=-1, \quad 
G_{0i}=\sum_{I}\epsilon(\epsilon g^2_{Ii} + 
g^1_{Ii} X^I)\Theta _\epsilon (X^I) 
\label{targetmetric}
\end{equation} 
where the suffix $0$ denotes temporal components.  In the limit
$\epsilon \rightarrow 0$, the leading--order terms are the ones
proportional to the $g^1_{Ii}$ bending couplings.  From now on we
restrict ourselves to these.

For simplicity, we again consider photons moving along the $x$
direction, as in the previous section, in which case the relevant
metric perturbations are
\begin{equation} 
   h_{0x} = \epsilon \sum_{I} g^1_{Ii} X^I \Theta _\epsilon (X^I) 
\label{pert2} 
\end{equation} 
To evaluate $\sigma_1^2$ in terms of the two--point function of
$h_{0x}$, we consider the null geodesic in the presence of the small
metric perturbations (\ref{pert2}). To leading order in the
bending/recoil couplings $g^1_{Ii}$, one has:
\begin{equation} 
\langle\sigma _1^2 \rangle \sim  L^2 \int _x^{x'} dy\;\int _x^{x'} dy' \;\langle h_{0x}(y,t)h_{0x}(y',t')\rangle
\label{metric}
\end{equation} 
In the case of D--brane recoil/bending, the computation of the quantum
average $\langle \dots \rangle$ may be made in the Liouville--string
approach described in~\cite{ellis98,ellis99}.  In this case, the
quantum average $\langle \dots \rangle$ is replaced by a world--sheet
correlator calculated with a world--sheet action deformed by
(\ref{logpair}).  It is clear from (\ref{pert2}) that the two--point
metric correlator appearing in (\ref{metric}) is just the $\langle
D_\epsilon D_\epsilon\rangle$ world--sheet recoil two--point function
described in (\ref{twopoint}).  The result is therefore
\begin{equation} 
    \Delta t _{\mathrm{recoil/bending}} \sim \frac{L}{c}\left
    ( \sum_{I}|g^1_{Ii}|^2\right)^{1/2}
\label{recoilbending}
\end{equation} 
To obtain an order of magnitude estimate of the effect, we take into
account the fact that, for $I=0$, the coupling $g^1_{0i} \sim U_i$ is
the recoil velocity of the world 3--brane. Viewed as a very massive
non--relativistic string soliton, in a dual string theory with
coupling $g_s$, the three--dimensional brane world would
have~\cite{ellis98,mavro+szabo} a recoil velocity $U_i \sim g_s E/M_s
$, where $M_s$ is the fundamental string scale, and $E$ is the typical
low--energy scale of the photon propagating on the brane.

Consistent embeddings of the picture of~\cite{dimopo98} into a
string--theoretical framework have been made
in~\cite{antoniadis98,antoniadis99}.  There are various string
theories which seem theoretically consistent with the picture
of~\cite{dimopo98}.  We now argue that the recoil expected in any
realistic model compatible with general relativity in the
$(3\!+\!1\!+\!n)$--dimensional space--time places strong restrictions
on such models. We examine two explicit cases, namely type I$'$ and
type II strings.  In the first case~\cite{antoniadis98}, D--3--brane
configurations appear to be consistent solutions of the model, but
with the restriction that only gravitational closed--string states can
propagate in the bulk, exactly as advocated in~\cite{dimopo98}.  In
this case, the string coupling \(g_s\) is given by the four
dimensional Yang--Mills gauge fine structure constant at the string
scale \(g_s = 4 \alpha _G,\) so that (\ref{recoilbending}) yields
$\Delta t \sim \alpha_G (L E)/ M_s$.  Since the astrophysical data on
GRBs and other sources are sensitive to $\Delta t \sim (L E)/ M_{QG}$
with $M_{QG} \sim 10^{15}$~GeV~\cite{amelino98}, it seems that, in
such a type I$'$ scenario, $M_s$ cannot lie in the TeV range as
originally proposed~\cite{dimopo98}.

In the case of type II closed strings, the picture of large extra
dimensions can be accommodated~\cite{antoniadis99} provided one uses
an extremely weak string coupling:
\begin{equation}
g_s \sim \alpha _G^{-1/2} \frac{M_s}{M_p^{(4)}} 
\label{pioline}
\end{equation} 
which is of order $10^{-14}$ for $M_s \sim {\rm TeV }$.  In such a
scenario there are D--brane solutions, in particular D--5
Neveu--Schwarz branes in which two of the longitudinal dimensions (as
well as the extra transverse dimensions) are assumed to be of the
string size.  This implies that the brane looks effectively
three--dimensional for low--energy physics, as in the type I$'$ string
case.  Within our recoil framework, couplings of the form
(\ref{pioline}) will lead to light--cone broadening
(\ref{recoilbending}) which depends solely on the four--dimensional
Planck scale,
\[\Delta t_{\mathrm{recoil/bending}} \sim \frac{L}{c} \,
\frac{E}{M_p^{(4)}}\] Comparing with the bounds set
in~\cite{amelino98}, we see that this possibility is not excluded, but
is also not far beyond the present experimental sensitivity.  We also
note that, in this latter framework, higher--order quantum phenomena,
due to higher world--sheet topologies~\cite{ellis98,ellis99}, are
suppressed by extra powers of the (weak) string coupling $g_s \sim
10^{-14}$ and hence are negligible.  It is possible to consider models
\cite{antoniadis99} with some of the transverse dimensions larger
which would imply higher values for the string coupling \(g_s\).  
For $M_s = 10$~TeV, the astrophysical data~\cite{amelino98} and our
recoil formalism place the phenomenological restriction $g_s \lesssim
10^{-11}$ on the maximum string coupling through the resulting
light--cone broadening effect (\ref{recoilbending}).

\section{Conclusions}

We have examined in this letter light--cone broadening effects in the
context of non--periodic extra dimensions between two parallel brane
worlds, one of which represents the observable universe.  We have
considered the phenomenon from both field-- and string--theoretical
viewpoints, by analysing the r\^{o}le of coherent graviton
fluctuations on the propagation of photons on one brane.

In the field--theoretical case, we have estimated that the light--cone
broadening scales with the distance $L$ traversed by the photons as
\[ \Delta t \sim \sqrt{L \; \ln L}.\]  For astrophysical sources
such as gamma--ray bursters, the order of the effect is about
\(10^{-12}\) seconds, which falls well below the sensitivity of
observations.  However, the sensitivity of gravity--wave
interferometer experiments is much better, and for these experiments
we estimate a light--cone broadening of the order of \(10^{-25}\)
seconds, which may well lie within their sensitivity.

In the string case, we have found that the dominant contributions to
the phenomenon come from the recoil of the D--brane due to the
scattering of closed--string states (gravitons) propagating in the
bulk.  The recoil distorts the space--time around the D--brane,
resulting in a mean--field effect which implies stochastic
fluctuations in the arrival time of photons of energy $E$ on the brane
of order
\[ \Delta t \sim g_s \,L \; \frac{E}{M_s}.\]  
Such phenomena place strong restrictions on string--theoretical models
of extra dimensions~\cite{antoniadis98,antoniadis99}, particularly
type I$'$ models.

\newpage

\section*{Acknowledgements}

The authors take pleasure in thanking I. Antoniadis for helpful
discussions.  A.C.{--}S. would like to thank the CERN Theory Division
for hospitality, and gratefully acknowledges financial support from
P.P.A.R.C. (U.K.) (studentship number 96314661), which made his visit
to CERN possible.  The work of N.E.M. is partially supported by
P.P.A.R.C. (U.K.) under an Advanced Fellowship.

\end{document}